\documentstyle[12pt]{article}
\title{Hidden Orders and RVB formation
of the Four-Leg Heisenberg Ladder Model}
\author{Yoshihiro Nishiyama,${}^{1}$
        Naomichi Hatano${}^{1,2}$ and Masuo Suzuki${}^1$\\
{\it ${}^1$Department of Physics, University of Tokyo}\\
{\it Hongo 7-3-1, Bunkyo-ku, Tokyo 113}\\
{\it ${}^2$Lyman Laboratory of Physics, Harvard University}\\
{\it Cambridge,
Massachusetts 02138, USA}}
\date{}

\begin{document}
\begin{normalsize}
\maketitle

\section*{Abstract}
The ground state of the four-chain Heisenberg ladder model
is numerically investigated.
Hidden-order correlations suitable for the system are introduced and
calculated with an emphasis on the spatially isotropic point,
where a corresponding material exists.
The existence of a long-range hidden correlation indicates formation of
a short-range RVB state in the case of the antiferromagnetic
inter-chain coupling.
A transition between the phase of the ferromagnetic inter-chain coupling
and that of the antiferromagnetic one is discussed.

\noindent KEYWORDS: hidden correlation, ladder model, resonating valence
bond, Haldane phase

\section{Introduction}
In recent years, considerable attention has been devoted to
ground-state properties of low-dimensional quantum systems.
The interest was greatly stimulated by the discovery of the
high-temperature superconductivity,\cite{??} and by subsequent studies
which revealed the importance of the CuO$_2$ plane of high-$T_{\rm c}$
materials.
However, complete understanding of the high-$T_{\rm c}$ mechanism
still appears to be beyond our reach.
A possible new approach to the high-$T_{\rm c}$ mechanism
was provided by a recent experiment\cite{Takano92} on the novel material
Sr$_{2n_{\rm leg}-2}$Cu$_{2n_{\rm leg}}$O$_{4n_{\rm leg}-2}$.
This material has periodic line defects in the CuO$_2$ plane, and thus
is composed of ladders with $n_{\rm leg}$ legs interacting weakly with each
other.
The limit $n_{\rm leg}\to\infty$ yields the high-$T_{\rm c}$
superconductors.

At half-filling this material may be described well by
the antiferromagnetic Heisenberg model defined on a ladder:
\begin{equation}
{\cal H}=J\sum_{l=1}^{n_{\rm leg}}\sum_{i=1}^{L}
\mbox{\boldmath $S$}_{l,i}\cdot\mbox{\boldmath $S$}_{l,i+1}
+J'\sum_{l=1}^{n_{\rm leg}-1}\sum_{i=1}^{L}
\mbox{\boldmath $S$}_{l,i}\cdot\mbox{\boldmath $S$}_{l+1,i}.
\label{Hamil}
\end{equation}
(Here $\mbox{\boldmath $S$}_{l,i}$ denotes the $S=1/2$ spin at the $i$th site
of
the $l$th chain. We put $J=1$ hereafter.)
It was reported,\cite{Dagotto92} indeed, that the double chain
($n_{\rm leg}=2$) with $J\sim J'$ develops the paring correlation upon doping.
This report was followed by the suggestion\cite{Rice93,Gopalan94,Sigrist94}
that a set of ladders interacting weakly with frustration might show
the superconductivity.
Thus, characterization of the ground state of the ladder models can be
an essential step to understanding the high-temperature
superconductivity.

It has been conjectured\cite{Rice93,Gopalan94,Sigrist94}
for $J=J'$ that the ground state of the model (\ref{Hamil})
with even $n_{\rm leg}$ is spin liquid with the energy gap,
while that with odd $n_{\rm leg}$ is critical, or gapless.
This conjecture is supported by a theorem,\cite{Affleck88a}
numerical calculations up to
$n_{\rm leg}=4$,\cite{Dagotto92,Parola93,White94}
and a scaling theory.\cite{Hatano95}
In order to explain this remarkable conjecture intuitively,
White {\it et al.}\cite{White94}\ proposed a resonating-valence-bond
(RVB) picture of the ground state of the antiferromagnetic ladder models.
This was followed by a proposal\cite{Nishiyama95,Watanabe95,White95}
of a hidden-order correlation for the RVB ground state of the double chain:
\begin{equation}
{\cal O}^z_{\rm RVB}(|i-j|)=
\left\langle (S^z_{1,i}+S^z_{2,i+1})
{\rm e}^{{\rm i}\pi\sum_{k=i}^{j-1}(S^z_{1,k}+S^z_{2,k+1})}
(S^z_{1,j}+S^z_{2,j+1}) \right\rangle.
\label{RVB_double}
\end{equation}
The existence of the long-range RVB correlation
supported\cite{Nishiyama95,Watanabe95,White95} the RVB
picture for $n_{\rm leg}=2$.

The present RVB correlation (\ref{RVB_double})
was also shown\cite{Nishiyama95} to be very useful
in discussing another issue, namely the criticality of the model
(\ref{Hamil}).
Suppose that we change the value of the coupling across the ladder,
$J'$ (the inter-chain coupling).
Then the above conjecture arises the following question.
The model is known to be critical, or gapless\cite{Bethe} for $J'=0$,
whereas the model has the gap for $J'=J$ according
to the above conjecture.
How does the energy gap emerge as $J'$ changes?
In the previous paper\cite{Nishiyama95} we presented numerical calculations
of the RVB correlations as well as another hidden-order correlation,
namely the string correlation:\cite{Takada92}
\begin{equation}
{\cal O}^z_{\rm string}(|i-j|)=
\left\langle (S^z_{1,i}+S^z_{2,i})
{\rm e}^{{\rm i}\pi\sum_{k=i}^{j-1}(S^z_{1,k}+S^z_{2,k})}
(S^z_{1,j}+S^z_{2,j}) \right\rangle
\label{string_double}
\end{equation}
We showed\cite{Nishiyama95}
quite clearly for $n_{\rm leg}=2$ that the energy gap appears
as $\Delta E \sim J'^\nu$ with $\nu=1$.
This conclusion is consistent with other
studies.\cite{Hatano95,Barnes93,Watanabe94,Totsuka95c}

Hidden correlations were first introduced to investigate the ground state of
the integer-$S$ antiferromagnetic Heisenberg chains.
The hidden correlations relevant to the model are the following
string correlations,%
\cite{den89,Girvin89,Tasaki91,Oshikawa92,Totsuka95,Nishiyama95b,Schollwoeck95}.
\begin{equation}
\langle S^\alpha_i {\rm e}^{{\rm i}\frac \pi S \sum_{k=i}^{j-1} S^\alpha_k}
        S^\alpha_j \rangle\ (\alpha=x,\ z),
\label{plain_string}
\end{equation}
where $\{ \mbox{\boldmath $S$}_i \}$ denote the $S=2$ spins.
The correlations successfully detected
the valence-bond-solid (VBS) structure%
\cite{AKLT87,AKLT88,Arovas88}
of the ground state of the integer-spin chains as well as their criticality%
.\cite{Tasaki91,Oshikawa92}

Now that we know the usefulness of the RVB correlation in the case
$n_{\rm leg}=2$, it is a challenging problem to generalize the
definition (\ref{RVB_double}) to higher $n_{\rm leg}$, and to
explore the ground state and the criticality of general $n_{\rm leg}$
ladder models.
Here we consider the ground state of
the ladder model with {\it four} legs.
The Hamiltonian is given by (\ref{Hamil}) with $n_{\rm leg}=4$.
We generalize the definitions (\ref{RVB_double}) and
(\ref{string_double}) for the four-leg ladder, and show
the RVB structure of the ground state in terms of the correlations.
We also confirm the prediction of the scaling theory\cite{Hatano95}
that the ground-state phase diagram consists of
two disordered phases covering $J'>0$ and $J'<0$, respectively,
and the critical point between them, $J'_{\rm c}=0$.
This phase diagram is the
same as in the two-leg case.\cite{Barnes93,Watanabe94,Totsuka95c}

The ground-state properties are known in some cases.
In the limit $J'\to\infty$, the model is decoupled to independent
four-spin rungs.
The ground state is given by the product of singlets formed on
the rungs.\cite{Reigrotzki94}
Therefore, the energy gap
is given by $\Delta E(J')\sim J'$ in this strong-coupling region.
At the isotropic point $J'=1$, which is of experimental
interest,\cite{Takano92}
the energy gap $\Delta E$ and the correlation length $\xi$ have
been numerically estimated as $\Delta E=0.190$ and
$\xi=5\sim6$.\cite{White94}
At $J'=0$ the model is reduced to four $S=\frac12$
Heisenberg antiferromagnetic spin chains.
Hence, it is critical as noted above.
In the limit $J'\to-\infty$, the model converges to the $S=2$ Heisenberg
antiferromagnetic chain, which is massive
according to the Haldane conjecture.\cite{Haldane83a,Haldane83b}
The Haldane conjecture for $S=2$ has been confirmed
numerically.\cite{Nishiyama95,Schollwoeck95,Hatano93,???,Deisz93}

The present paper is organized as follows.
We introduce hidden correlations in the next section, and
present a naive discussion on the existence of the hidden correlations
in terms of the RVB argument.
In section 3, we show numerical results of the hidden correlations
with an emphasis on the case $J'=1$, where
the corresponding material is available.
We also present an argument on the existence of the RVB structure
by considering two double-chain ladders coupled weakly.
This argument intuitively shows the nature of the RVB ground state.
In section 4,
we elaborate on the criticality of the model in terms of the hidden
correlations.
In the last section, we give a summary of the present paper.

\section{RVB state and hidden correlations}
In this section, we introduce the hidden correlations for the
present four-leg-ladder model (\ref{Hamil}).
They are defined so that they can detect expected RVB patterns.

We present  schematic drawings of the expected
RVB patterns in Fig. 1 (a) for $J'>0$
and (b) for $J'<0$, respectively.
Each RVB pattern is arranged as follows.
First for $J'>0$,
White {\it et al.}\cite{White94} proposed that the ground state
is dominated by the RVB pattern with
vertical singlets and
horizontal singlets.
The state is stabilized by the resonance between the configuration of two
adjacent vertical singlets and the configuration of two horizontal singlets;
 see Fig. 1 (a).
This RVB picture was reported\cite{White94} to be very useful for understanding
several features of the ladder models.
Second, the expected RVB pattern for $J'<0$ is given as follows.
In the limit $J'\to-\infty$, the system (\ref{Hamil})
with $n_{\rm leg}=4$
converges to the $S=2$ Heisenberg antiferromagnetic chain.
The ground state of the chain was described\cite{Oshikawa92,Totsuka95,%
Nishiyama95b} well
in terms of the $S=2$ valence-bond-solid  state.
This VBS state is constructed in the following manner:
Suppose that each spin with $S=2$ consists of four spins with $S=\frac12$;
Form a singlet using two $S=\frac12$ spin of neighbouring sites;
Arrange the singlets so that each bond may have two singlets.
All the $S=\frac12$ spins are thus connected each other with the short-%
range valence bonds consequently.
We expect this state in the limit $J'\to-\infty$.
The resonance may arise among the different pairing patterns
for finite and negative $J'$.
Hence we have drawn the pattern which is  shown in Fig. 1 (b).
We expect that this pattern dominates
in the whole region $J'<0$.

In order to detect the above RVB patterns,
we define the following hidden
correlations:
\begin{eqnarray}
{\cal O}^z_{\rm RVB}(\theta,|i-j|)&=&
\left\langle (S^z_{1,i}+S^z_{2,i+1}+S^z_{3,i}+S^z_{4,i+1})
{\rm e}^{{\rm i}\theta\sum_{k=i}^{j-1}
(S^z_{1,k}+S^z_{2,k+1}+S^z_{3,k}+S^z_{4,k+1})} \right. \nonumber \\
& &\left. (S^z_{1,j}+S^z_{2,j+1}+S^z_{3,j}+S^z_{4,j+1}) \right\rangle
\label{RVB}
\end{eqnarray}
and
\begin{eqnarray}
{\cal O}^z_{\rm string}(\theta,|i-j|)&=&
\left\langle (S^z_{1,i}+S^z_{2,i}+S^z_{3,i}+S^z_{4,i})
{\rm e}^{{\rm i}\theta\sum_{k=i}^{j-1}%
(S^z_{1,k}+S^z_{2,k}+S^z_{3,k}+S^z_{4,k})} \right. \nonumber \\
& &\left. (S^z_{1,j}+S^z_{2,j}+S^z_{3,j}+S^z_{4,j}) \right\rangle.
\label{string}
\end{eqnarray}
We here generalized
the hidden correlations (\ref{RVB_double}) and (\ref{string_double}),
which have been
applied to the
double-chain ladder model.\cite{Nishiyama95,Watanabe95,White95}
The angle $\pi$ appearing in
(\ref{RVB_double})
and
(\ref{string_double})
is replaced by $\theta$ in the present definitions
(\ref{RVB})
and
(\ref{string}).

We show in the present paper that
the choice $\theta=\frac\pi2$ is the most relevant to
the four-chain ladder model.
We also show
that
the correlation (\ref{RVB}) develops in the phase $J'>0$
while
the corelation
(\ref{string}) develops in the phase $J'<0$.
Let us explain the reasons of these facts briefly, before going into details
in \S 3 and 4

First, in the limit $J'\to-\infty$,
the model converges to the $S=2$ Heisenberg antiferromagnetic
chain,
as is explained in the previous section.
The correlation (\ref{string}) then is reduced to
\begin{equation}
\langle S^z_i {\rm e}^{{\rm i}\theta\sum_{k=i}^{j-1} S^z_k} S^z_j \rangle,
\end{equation}
where $\{ {\bf S}_i \}$ denote the $S=2$ spins.
This correlation with $\theta=\frac\pi2$ is reported to remain
finite in the limit $|i-j|\to\infty$,\cite{Nishiyama95b,Schollwoeck95}
as was mentioned in \S1.
The correlation (\ref{RVB}), on the other hand, remains finite
in the limit $J'\to\infty$; see the next section.

Second, as was done for the double-chain ladder model,\cite{Nishiyama95}
we can show the existence of hidden orders schematically as follows.
Consider any spin configurations of the four-leg ladder,
satisfying the RVB pattern of Fig. 1 (a).
We present an example in Fig. 2 (a).
We notice that the set of the numbers
$\{ \tilde{S}_i^z\equiv S^z_{1,i}+S^z_{2,i}+S^z_{3,i}+S^z_{4,i}\}$ in
Fig. 2 (a)
satisfies the condition
\begin{equation}
\left|
\sum_{k=i}^{j} \tilde{S}^z_k
\right|\le2.
\label{condition}
\end{equation}
for ${}^\forall i$, ${}^\forall j$.
This is due to the formation of two singlets over two neighboring composite
spins
$\{ \tilde{\bf S}_i \}$.
It is easy to see that the quantity
\begin{equation}
\tilde{S}^z_i {\rm e}^{{\rm i}\frac\pi2 \sum_{k=i}^{j-1} \tilde{S}_k^z}
 \tilde{S}^z_j
\end{equation}
remains finite for any $i$ and $j$, when it is averaged over all configurations
that satisfy the condition (\ref{condition}).\cite{%
Oshikawa92,Totsuka95}
This explains that the RVB correlation (\ref{RVB}) remains finite
if the RVB configurations as Fig. 1 (a) are dominant in the ground state.
The above argument applies to the RVB patterns as Fig. 1 (b) if we
define the composite spin as
$\{ \tilde{S}^z_i\equiv S^z_{1,i}+S^z_{2,i+1}+S^z_{3,i}+S^z_{4,i+1}\}$;
see Fig. 2 (b).
We can expect that local destruction of the RVB patterns does not result
in vanishing of the long-range correlation.%
\cite{Nishiyama95}.


\section{RVB formation and the presence of the hidden order
for $J'>0$}
In this section,we investigate the ground state of the four-leg ladder
by means of the hidden correlations
(\ref{RVB}) and (\ref{string}) numerically.
The RVB picture explained in the above section
is confirmed.
We show that the choice $\theta=\frac\pi2$ is the most relevant to the
present system.

\subsection{Hidden correlations in the limit $J'\to\infty$}
In this subsection, we concentrate on the four-leg ladder
in the limit $J'\to\infty$.
The limiting point is expected to be the fixed point of the phase $J'>0$%
.\cite{Hatano95}.
Hence, characteristics at the point may be
relevant to those of the
whole region $J'>0$.

In this limit $J'\to\infty$, the Hamiltonian is reduced to
a set of independent rungs.
The ground state is given by the direct product of singlets that are formed
on the rungs.
We can calculate the long-range limit of  the hidden correlations
(\ref{RVB}) and (\ref{string}) exactly.
Though the ground state is spin liquid.%
\cite{White94} the hidden correlation
(\ref{RVB}) is long-ranged.
The correlation ${\cal O}^z_{\rm RVB}(\theta,|i-j|\to\infty)$
is plotted against
$\theta$ in Fig. 3.
The correlation has the maximum around $\theta=\pi/2$,
while the correlations at $\theta=0$ and $\pi$ are both vanishing.
These behaviors are intrinsic to the whole region of the phase $J'>0$
as we see below.
It is apparent, on the other hand, that the correlation (\ref{string})
is of short range, ${\cal O}^z_{\rm string}(\theta,|i-j|\to\infty)=0$.

\subsection{The hidden orders at $J'=1$}
In this subsection,
we concentrate on the system at $J'=1$, where the interaction is
spatially isotropic.
This isotropic system is of experimental interest.%
\cite{Takano92}.
A strong-inter-chain-coupling expansion starts to fail
at this point.\cite{Reigrotzki94}
We employed the density-matrix renormalization-group
method \cite{White93,White94b}
in order to treat large systems approximately.
We conclude that ground state properties are indeed consistent with
the RVB picture given in the previous section
and the recent proposal that the point $J'\to\infty$ is the fixed point
of the phase $J'>0$.\cite{Hatano95}

First of all, we show the precision of the present renormalization-group
calculation.
In Fig. 4 we plotted the relative error of the ground state energy with $J'=1$
and $L=6$
against the approximate level $m$.
The parameter $m$ is the number of states kept;\cite{White93,White94b}
in this method,
we treat only $m$ states in the course of the renormalization.
It should be noted that the precision of the correlations is worse
than that of the ground-state energy.
The reason may be as follows.
Because the Hilbert space is restricted to the $m$ states,
this renormalization-group
method may be regarded as a kind of variational method.
In many cases, the variational calculation yields worse estimation
for the correlation functions.

The correlations ${\cal O}^z_{\rm RVB}(\theta,21)$
and ${\cal O}^z_{\rm string}(\theta,21)$ are plotted in Fig. 5 (a) and (b),
respectively.
The maxima of the correlations are located around $\theta=\frac\pi2$.
This is also reported in the previous subsection for the
system with $J'\to\infty$.
We hence observe that the most
relevant angle for the correlations is given by $\theta=\pi/2$
as is expected from the discussions in \S 2.
In order to estimate the infinite distance limit of the correlations,
we plotted ${\cal O}^z_{\rm RVB}(\frac\pi2,|i-j|)$
and ${\cal O}^z_{\rm string}(\frac\pi2,|i-j|)$ against $1/|i-j|$
in Fig. 6 (a) and (b), respectively.
The correlation (\ref{RVB}) remains finite;
\begin{equation}
{\cal O}^z_{\rm RVB}\left(\frac\pi2,|i-j|\to\infty\right)\approx0.65.
\end{equation}
(The correlation would have developed fully as ${\cal O}^z(\frac\pi2)=1$,
if the ground state satisfied the condition (\ref{condition}) completely.)
On the other hand, it is seen that the correlation (\ref{string})
is very small.
Considering the precision of the numerical calculators,
we conclude that the correlation (\ref{string}) is of short range.

Finally, we show the results of the
correlation (\ref{RVB}) with the angle $\theta=0$.
Note that this is nothing but the N\'eel correlation.
We present a semi-logarithmic plot of the correlation against $|i-j|$;
see Fig. 7.
It is seen that it does decrease exponentially.
The correlation length is somewhat consistent with an estimate
in Ref. \cite{White94}.

\subsection{Mechanism of the RVB order with $\theta=\frac\pi2$}
In this subsection,
we clarify the essential mechanism of the development
of the correlation (\ref{RVB}) with $\theta=\frac\pi2$
in the region $J'>0$.

For this purpose, let us decouple
the four-leg ladder system into two ladder systems.
The Hamiltonian is given by
\begin{eqnarray}
{\cal H}&=&\sum_{l=1}^{4}\sum_{i=1}^{L}
\mbox{\boldmath $S$}_{l,i}\cdot\mbox{\boldmath $S$}_{l,i+1} \nonumber \\
& &+J'\sum_{i=1}^{L}
\mbox{\boldmath $S$}_{1,i}\cdot\mbox{\boldmath $S$}_{2,i} \nonumber \\
& &+J''\sum_{i=1}^{L}
\mbox{\boldmath $S$}_{2,i}\cdot\mbox{\boldmath $S$}_{3,i} \nonumber \\
& &+J'\sum_{i=1}^{L}
\mbox{\boldmath $S$}_{3,i}\cdot\mbox{\boldmath $S$}_{4,i}.
\label{inter_ladder}
\end{eqnarray}
The effect of the inter-ladder coupling $J''$ is analysed with the aid of
numerical
simulations below.
Reigrotzki {\it et al.}\cite{Reigrotzki94}
suggested that the decoupled system ($J''=0$) is a convenient starting
point for
investigating the four-leg ladder.

First, we consider the completely decoupled case $J''=0$.
The hidden correlation (\ref{RVB})
is reduced to
\begin{eqnarray}
{\cal O}^z_{\rm RVB}(\theta,|i-j|)&=&
2\langle \sigma_i^z
       {\rm e}^{{\rm i} \theta \sum_{k=i}^{j-1} \sigma_k^z}
         \sigma_j^z \rangle
\langle
       {\rm e}^{{\rm i} \theta \sum_{k=i}^{j-1} \tau_k^z}
       \rangle\nonumber\\
& &+2 \langle
       {\rm e}^{{\rm i} \theta \sum_{k=i+1}^{j} \sigma_k^z}\sigma_j^z
         \rangle
    \langle
       {\rm e}^{{\rm i} \theta \sum_{k=i}^{j-1} \tau_k^z}\tau_j^z
         \rangle,
\label{decoupled_RVB}
\end{eqnarray}
where $\sigma_i^z=S_{1,i}^z+S_{2,i+1}^z$ and
$\tau_i^z=S_{3,i}^z+S_{4,i+1}$.
The brackets here denote the ground-state expectation
value of a single two-leg ladder system.

In Fig. 8, we show an example of spin
configurations $\{ \sigma_i^z  \}$ and $\{ \tau_i^z \}$
for each single-ladder system.
The configurations are expressed in terms of the height of steps.
The upward (downward) step at the position $i$ of the upper drawing stands for
$\sigma^z_i=1$ ($\sigma^z_i=-1$).
The same rule applies to the lower drawing.
The example here satisfies the conditions
\begin{equation}
\left| \sum_{k=i}^j \sigma^z_k \right| \le 1\
{\rm and}\
\left| \sum_{k=i}^j \tau^z_k \right| \le 1
\label{conditions}
\end{equation}
for ${}^\forall i$, ${}^\forall j$;
therefore the sites with
the magnetization $1$ and $-1$ appear alternately with
the sites with $0$ being inserted between them.
Indeed, such configurations are expected to dominate the ground state
of the two-leg ladder system
\cite{Nishiyama95,%
Watanabe95,White95},
according to the RVB theory for the two-leg ladder.\cite{White94}
RVB pattern as in Fig. 9 satisfies the condition (\ref{conditions}).
We calculated the two terms in eq. (\ref{decoupled_RVB})
for the exemplified
configuration in Fig. 8 fixing the angle $\theta$ to $\theta=\frac\pi2$.
The expectation values indicated in Fig. 8 are the average over
the depicted configuration
and the
reflected ($\sigma_i^z\to-\sigma_i^z$, $\tau_i^z\to-\tau_i^z$)
one.
We can see that the correlation (\ref{decoupled_RVB})
does not vanish only at the sites
where the steps of the upper and lower
ladders synchronize coincidently.

We can also show after a
similar analysis
that the hidden correlation (\ref{RVB})
does not develop for $\theta=0$ and $\theta=\pi$,
if we restrict the configurations with the condition (\ref{conditions}).
This is the same as what we observe in the previous subsection.
Thus, the analysis based on the decoupled system appears to be
fairly relevant.

Now we investigate the effect of the inter-ladder coupling in  terms
of
the above RVB picture numerically.
As a consequence, we present more detailed information on the RVB pattern.
In Fig. 10, we show the hidden correlation (\ref{RVB}) for the system
with $L=6$, $J'=1$ and $J''$ varied.
We observe that the inter-ladder coupling $J''$
stabilizes the hidden correlation.
Recall that the  essential mechanism of the development
of the RVB correlation is the synchronization of the configurations
of the upper-half ladder
and the lower-half ladder; see Fig. 8.
The RVB correlation ${\cal O}^z_{\rm RVB}$ for $J''=0$ expresses the
contribution of the coincidental synchronization.
We can observe from Fig. 10 that the inter-ladder coupling enhances the
synchronization.

We can explain the reason of the enhancement as follows.
According to the RVB picture, the ground state is stabilized by the
resonance between the configuration of two vertical singlets and that of
two horizontal singlets.\cite{White94,Nishiyama95}
When the inter-ladder coupling $J''$ is turned on,
the resonance
shown in Fig. 11 becomes possible.
To take advantage of the resonance, singlets along the upper-half ladder and
those
along the lower-half ladder may tend to appear at the same position
$i$.
This effect can enhance the synchronization of the configurations.

\section{Criticality and the phase diagram}
In this section, we investigate the development of the hidden correlations
(\ref{RVB}) and (\ref{string})
and their criticality when we change the parameter $J'$.
We show the results of the
exact-diagonalization method under the periodic-boundary
condition.
The result is consistent with that of the scaling theory\cite{Hatano95}.

We plotted the values of ${\cal O}^z_{\rm RVB}(\frac\pi2,\frac L2)$
and ${\cal O}^z_{\rm string}(\frac\pi2,\frac L2)$ for the system of the size
$L=4$ and $6$ in Fig. 12.
As is expected,
we observe that the correlation ${\cal O}^z_{\rm RVB}(\frac\pi2)$
develops in the  region $J'>0$, while the correlation
${\cal O}^z_{\rm string}(\frac\pi2)$ develops in the region $J'<0$.
We readily know the values of the correlations in two limiting cases.
In the limit $J'\to\infty$, we showed
${\cal O}^z_{\rm RVB}(\frac\pi2)=0.386895$.
In the limit $J'\to-\infty$,
the corelation (\ref{string}) corresponds to
the string correlation of the $S=2$ Heisenberg chain.
It is reported%
\cite{Nishiyama95b,Schollwoeck95}
that we have ${\cal O}^z_{\rm string}(\frac\pi2)\approx0.65$.

In order to see the development of the correlations and the criticality,
we calculated the  squared order parameters,
\begin{equation}
\langle O^\dagger_{\rm RVB} (\frac\pi2)
O_{\rm RVB} (\frac\pi2)\rangle,
\end{equation}
and
\begin{equation}
\langle O^\dagger_{\rm string}(\frac\pi2)
O_{\rm string} (\frac\pi2)\rangle,
\end{equation}
where
\begin{equation}
\label{binRVB}
O_{\rm RVB}(\theta)=\frac1L\sum_{i=1}^L
{\rm e}^{{\rm i}\theta\sum_{k=1}^{i-1}
(S^z_{1,k}+S^z_{2,k+1}+S^z_{3,k}+S^z_{4,k+1})}
(S^z_{1,i}+S^z_{2,i+1}+S^z_{3,i}+S^z_{4,i+1})
\end{equation}
and
\begin{equation}
\label{binstring}
O_{\rm string}(\theta)=\frac1L\sum_{i=1}^L
{\rm e}^{{\rm i}\theta\sum_{k=1}^{i-1}
(S^z_{1,k}+S^z_{2,k}+S^z_{3,k}+S^z_{4,k})}
(S^z_{1,i}+S^z_{2,i}+S^z_{3,i}+S^z_{4,i}).
\end{equation}
We plotted the results
in Fig. 13 (a) and (b), scaling the data
by the factor $L^{\frac14}$.
This is because of the following reason.
At the critical point $J'_{\rm c}=0$, the correlations are reduced to
\begin{eqnarray}\label{square}
{\cal O}^z_\alpha(\theta,r)&=&
\left({\cal O}^z_{S=1/2}(\theta,r)\right)^4\quad(\alpha={\rm string,\ RVB}) \\
 &\sim&r^{-\eta_\alpha(\theta)}, \nonumber
\end{eqnarray}
where ${\cal O}^z_{S=1/2}$ denotes the string correlation for the
$S=1/2$ Heisenberg chain:\cite{Hida92}
\begin{equation}\label{halfstring}
{\cal O}^z_{S=1/2}(\theta,|i-j|)
=\left\langle S^z_i{\rm e}^{{\rm i}\theta
\sum_{k=i}^{j-1}S_k^z}S^z_j\right\rangle.
\end{equation}
If we know the correlation exponent of (\ref{halfstring}) defined in
\begin{equation}
{\cal O}^z_{S=1/2}(\theta,r)\sim r^{-\eta_{S=1/2}(\theta)},
\end{equation}
the relation (\ref{square}) is immediately followed by
\begin{equation}\label{eta2}
\eta_\alpha(\theta)=4\eta_{S=1/2}(\theta)
\end{equation}
both for $\alpha={\rm string}$ and $\alpha={\rm RVB}$.
Generalizing Hida's analysis\cite{Hida92}
we obtain the formula $\eta_{S=1/2}(\theta)=\frac14(\frac\theta\pi)^2$.
Therefore, the exponent is given by
$\eta_\alpha(\frac\pi2)=\frac14$.

The scaled order parameters are invariant at the critical point
for various system sizes.
In Fig. 13 (a) and (b), we observe the crossing points close to the
expected critical point $J'_{\rm c}=0$.\cite{Hatano95}
(Another crossing point which appears in Fig. 13 (b)
around $J'\approx1.5$ may have arose owing to
the limited system size.)
The hidden correlations appear to obey the behavior explained in  \S2.

\section{Summary}
The ground state of the four-chain ladder model with both ferromagnetic
and antiferromagnetic inter-chain coupling $J'$ has been analysed
by means of the hidden correlations (\ref{RVB}) and (\ref{string}).
The hidden correlations can indicate the development of
the corresponding RVB states.

The hidden correlation (\ref{RVB}) develops in the phase
$J'>0$, while the hidden correlation (\ref{string}) develops in the phase
$J'<0$.
Though both the two phases are disordered,
they are characterized in terms of the  hidden long-range
orders.
It indicates that the RVB structure varies drastically at the critical point
$J'_{\rm c}=0$.
These results are consistent with a recent proposal\cite{Hatano95} that
the point $J'\to\infty$ $(-\infty)$
is the fixed point of the region $J'>0$ $(<0)$.
A detailed analysis of
RVB pattern has been reported for the phase $J'>0$.
It implies that valence bonds formed along the ladder are
not negligible as are in the case of the double-chain ladder
model.\cite{Nishiyama95}
This tendency may be more apparent as the number of the legs is increased.

We are in position to conclude
that, in the phase $J'>0$,
the ground state of the ladder models with two and four legs
is the RVB state proposed by White {\it et al.}\cite{White94}.
The RVB pattern
may be common
to the arbitrary ladder models with
even legs.

\section*{Acknowledgement}
Our computer programs
are partly based on the subroutine package "TITPACK Ver. 2"
coded by Professor H. Nishimori.

The authors thank T. Momoi, A. Terai and K. Totsuka for valuable comments.

{\bf Figure captions}

Fig. 1 Schematic drawings of the expected RVB states.
The RVB pattern of the type
(a) ((b)) is dominant in the phase $J'>0$ ($J'<0$).

Fig. 2 Examples of spin configurations.
The configurations (a) and (b) are arranged so as to satisfy the RVB
patterns
in Fig. 1 (a) and (b), respectively.
The composite magnetization $\tilde{S^z_{i}}$
is shown below each configurations.
The full string order develops in both cases.

Fig. 3 The long-range hidden correlation
${\cal O}^z_{\rm RVB}(\theta,|i-j|\to\infty)$ against the angle $\theta$
at $J'\to\infty$.

Fig. 4 The relative error of the ground-state energy that is calculated by
means
of the density-matrix renormalization-group method
for the system with the parameter $J'=1$ and the system size $L=6$
under the open-boundary condition.

Fig. 5 The hidden correlations
(a) ${\cal O}^z_{\rm RVB}(\theta,21)$ and
(b) ${\cal O}^z_{\rm string}(\theta,21)$
with the angle $\theta$
varied.
The maxima are located around $\theta=\frac\pi2$.

Fig. 6 The hidden correlations
(a) ${\cal O}^z_{\rm RVB}(\frac\pi2,|i-j|)$ and
(b) ${\cal O}^z_{\rm string}(\frac\pi2,|i-j|)$
plotted against the inverse
of the distance $1/|i-j|$.

Fig. 7  A semi-logarithmic plot of ${\cal O}^z_{\rm RVB}(0,|i-j|)$
against $|i-j|$.
For reference, we show the expected long-range behavior
$\propto {\rm e}^{-|i-j|/5.5}$\cite{White94} as the broken line.

Fig. 8 An example of the
spin configuration for the upper double-chain ladder and
the lower ladder. Two terms that appear in eq. (\ref{decoupled_RVB})
are calculated for the configurations, respectively.
The expectation values are the average over the depicted configuration and
the reflected one ($\sigma_i^z\to-\sigma_i^z$, $\tau_i^z\to-\tau_i^z$).

Fig. 9 Example of the expected RVB pattern for the decoupled ($J''=0$)
system.

Fig. 10
The hidden correlation
${\cal O}^z_{\rm RVB}(\frac\pi2,\frac L2)$
for the system with $L=6$, $J'=1$ and $J''$ varied.

Fig. 11 As the inter-ladder coupling $J''$ is turned on,
the resonance
between the upper and lower ladders appears.

Fig. 12 The $J'$ dependence of the hidden correlations
${\cal O}^z_{\rm RVB}(\frac\pi2,\frac L2)$
and
${\cal O}^z_{\rm string}(\frac\pi2,\frac L2)$.

Fig. 13 The $J'$ dependance of the scaled hidden orders:
(a) $L^{\frac14}\langle O^\dagger_{\rm RVB}(\frac\pi2)O_{\rm
RVB}(\frac\pi2)\rangle$
and (b) $L^{\frac14}\langle O^\dagger_{\rm string}(\frac\pi2)O_{\rm
string}(\frac\pi2)\rangle$.
The intersection point may indicate the critical point.

\end{normalsize}
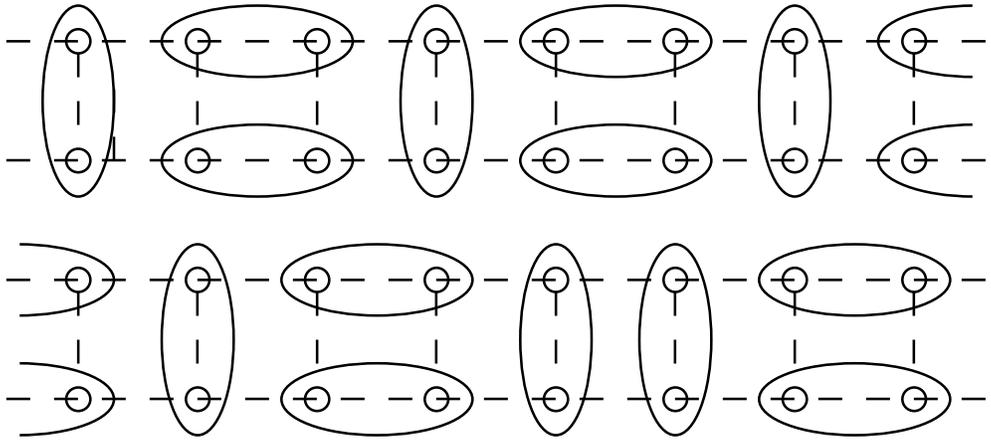
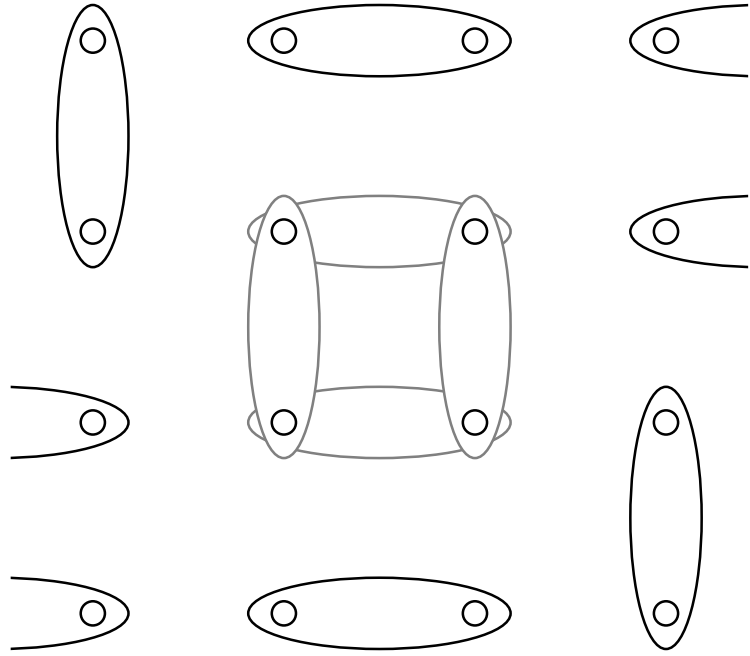
\end{document}